\documentclass{spie}  

\usepackage{amsmath,amsfonts,amssymb}
\usepackage{graphicx}
\usepackage[colorlinks=true, allcolors=blue]{hyperref}

\usepackage{physics}

\title{Prototype Development and Calibration of the CUbesat Solar Polarimeter (CUSP)}

\author[a]{Nicolas De Angelis}
\author[a]{Abhay Kumar}
\author[a]{Sergio Fabiani}
\author[a]{Ettore Del Monte}
\author[a]{Enrico Costa}
\author[a,b]{Giovanni Lombardi}
\author[a]{Paolo Soffitta}
\author[a,c]{Andrea~Alimenti}
\author[d,e]{Riccardo~Campana}
\author[f]{Mauro~Centrone}
\author[d]{Giovanni~De~Cesare}
\author[a]{Sergio~Di~Cosimo}
\author[a]{Giuseppe~Di~Persio}
\author[a]{Alessandro~Lacerenza}
\author[a]{Pasqualino~Loffredo}
\author[g]{Gabriele~Minervini}
\author[a]{Fabio~Muleri}
\author[h]{Paolo~Romano}
\author[a]{Alda~Rubini}
\author[a]{Emanuele~Scalise}
\author[a,c]{Enrico~Silva}
\author[i]{Davide~Albanesi}
\author[j]{Ilaria~Baffo}
\author[k]{Daniele~Brienza}
\author[i]{Valerio~Campamaggiore}
\author[l]{Giovanni~Cucinella}
\author[m]{Andrea~Curatolo}
\author[i]{Giulia~de~Iulis}
\author[i]{Andrea~Del~Re}
\author[l]{Vito~Di~Bari}
\author[l]{Simone~Di~Filippo}
\author[k]{Immacolata~Donnarumma}
\author[j]{Pierluigi~Fanelli}
\author[n]{Nicolas~Gagliardi}
\author[i]{Paolo~Leonetti}
\author[k]{Matteo~Mergè}
\author[m,n]{Dario~Modenini}
\author[l]{Andrea~Negri}
\author[m]{Daniele~Pecorella}
\author[l]{Massimo~Perelli}
\author[n]{Alice~Ponti}
\author[i]{Francesca~Sbop}
\author[m,n]{Paolo~Tortora}
\author[k]{Alessandro~Turchi}
\author[k]{Valerio~Vagelli}
\author[k]{Emanuele~Zaccagnino}
\author[i]{Alessandro~Zambardi}
\author[j]{Costantino~Zazza}

\affil[a]{INAF-IAPS\\ via del Fosso del Cavaliere 100, 00133 Rome, Italy}
\affil[b]{Department of Enterprise Engineering ”Mario Lucenti”, University of Rome "Tor Vergata", Via Cracovia 50, 00133 Rome, Italy}
\affil[c]{Department of Industrial, Electronic and Mechanical Engineering, "Roma Tre" University, via V. Volterra 62, 00146 Rome, Italy}
\affil[d]{INAF-OAS Bologna\\ via Gobetti 93/3, 40129 Bologna, Italy}
\affil[e]{INFN Sezione di Bologna, viale Berti Pichat 6/2, 40127 Bologna, Italy}
\affil[f]{INAF-OAR\\ via Frascati 33, 00040, Monte Porzio Catone, Italy}
\affil[g]{INAF-Headquarters\\ viale del Parco Mellini 84, 00136 Rome, Italy}
\affil[h]{INAF-OACT\\ Via S. Sofia 78, 95123, Catania, Italy}
\affil[i]{DEDA Connect s.r.l.\\ via Vincenzo Lamaro 51, 00173 Rome, Italy}
\affil[j]{DEIM, University of "La Tuscia", Largo dell’Università, 01100 Viterbo, Italy}
\affil[k]{ASI, via del Politecnico snc\\ 00133 Rome, Italy}
\affil[l]{IMT s.r.l.\\ via Carlo Bartolomeo Piazza 30, 00161 Rome, Italy}
\affil[m]{Department of Industrial Engineering - Alma Mater Studiorum Università di Bologna, Via Montaspro 97, 47121 Forlì, Italy}
\affil[n]{Interdepartmental Centre for Industrial Aerospace Research - Alma Mater Studiorum Università di Bologna, Via Carnaccini 12, 47121 Forlì, Italy}

\authorinfo{Further author information: (Send correspondence to N.D.A.)\\N.D.A.: E-mail: \href{mailto:nicolas.deangelis@inaf.it}{nicolas.deangelis@inaf.it}}

\begin{document} 
\maketitle

\begin{abstract}
The space-based CUbesat Solar Polarimeter (CUSP) mission aims to measure the linear polarization of solar flares in the hard X-ray band by means of a Compton scattering polarimeter. CUSP will allow to study the magnetic reconnection and particle acceleration in the flaring magnetic structures of our star with its unprecedented sensitivity to solar flare polarization. CUSP is a project in the framework of the Alcor Program of the Italian Space Agency aimed to develop new CubeSat missions. It has been proposed as a constellation of a two Cubesat mission to monitor the Sun for Space Weather, and will proceed with a single-satellite asset in its baseline implementation.

In the frame of CUSP's Phase B study, that started in December 2024 for a 1-year period, we present the development status of this dual-phase polarimeter. Preliminary laboratory results using two chains of acquisition will be discussed. The first chain of acquisition, based on the Hamamatsu R7600 multi-anode photomultiplier tubes coupled to plastic scintillator bars and read out by the MAROC-3A ASIC, is used to detect the Compton scattering of incoming photons. On the other hand, GAGG crystals coupled to avalanche photo-diodes with a readout based on the SKIROC-2A ASIC are used to absorb the scattered photons. By reconstructing the azimuthal scattering direction for many incoming photons, one can infer the linear polarization degree and angle of the source. We will discuss the calibration results obtained with our prototype detector by using well-known radioactive isotopes, allowing us to assess the performances of our detector over the full 25-100 keV energy range.
\end{abstract}


\section{INTRODUCTION: SOLAR FLARES AND POLARIZATION}\label{sec:intro}

As the only star of our solar system and the dominant driver of the heliospheric environment, the Sun exhibits a rich variety of dynamic and eruptive phenomena. Among the most energetic of these are solar flares, which result from the rapid release of magnetic energy stored in the solar corona. Flares often occur in conjunction with coronal mass ejections (CMEs) -- massive expulsions of plasma and magnetic field into interplanetary space. Together, solar flares and CMEs are the primary engines of space weather, with the potential to affect satellite operations, radio communications, GPS systems, power grids, and even astronaut safety.

Flares are thought to be triggered by magnetic reconnection processes, wherein stressed magnetic field lines in the corona rearrange and reconnect, converting magnetic energy into particle acceleration, plasma heating, and electromagnetic radiation across a broad spectral range. During a flare, accelerated electrons interact with the dense chromosphere and produce X-ray bremsstrahlung as they decelerate. This X-ray emission consists of both thermal and non-thermal components, with the latter dominating during the impulsive phase in the hard X-ray band ($\geq$25 keV) \cite{Temmer2016, Nagasawa2022}. Non-thermal X-rays carry key information about the distribution, anisotropy, and beaming of accelerated particles, offering a direct probe into the microphysics of reconnection and particle transport \cite{Zharkova2010}.

Despite extensive multi-wavelength observations, major open questions remain in flare physics: What is the exact geometry and location of the reconnection site? How are particles accelerated so efficiently? What determines the directionality and pitch-angle distribution of high-energy electrons? While spectroscopic and imaging observations have provided important constraints, they are often insufficient to break degeneracies between competing theoretical models.

X-ray polarimetry adds a powerful, complementary diagnostic. The degree and orientation of linear polarization in the hard X-ray emission depends sensitively on the geometry of the magnetic field, the directionality of electron beams, and the viewing angle\cite{Zharkova2010, Jeffrey2020}. In particular, non-thermal bremsstrahlung from beamed electrons is expected to be strongly polarized (tens of percent), while thermal bremsstrahlung is only weakly polarized. Therefore, polarimetric measurements in the 25–100 keV band can effectively distinguish between thermal and non-thermal emission mechanisms, probe electron anisotropy, and place direct constraints on the flare magnetic geometry.

To date, X-ray polarization measurements of solar flares have been scarce and statistically limited, often yielding only upper limits or marginal detections \cite{Tindo1970, Tindo1972a, Tindo1972b, Tramiel1984, Boggs2006, SuarezGarcia2006}. Advancing this field requires dedicated instruments with sufficient sensitivity and time resolution to resolve the impulsive dynamics of flares. Given the implications for both fundamental heliophysics and applied space weather forecasting, high-significance solar flare polarimetry remains a high-priority but under-explored frontier in solar physics. The CUSP mission aims to address this gap by delivering dedicated, high-sensitivity measurements of solar flare X-ray polarization in the 25–100 keV range, enabling time-resolved studies of flare dynamics and providing critical insights into particle acceleration and magnetic field geometries that drive solar eruptive events.

After presenting CUSP's polarimeter design and working principle, we report here the first laboratory measurements performed using single-channel detectors and development boards, aimed at testing both the basic functionality and preliminary performance of the detection system.

\newpage
\section{SOLAR FLARE POLARIMETRY WITH CUSP}\label{sec:sf_cusp}

In order to measure the polarization of solar flares in the hard X-ray band, the CUbesat Solar Polarimeter (CUSP) mission is under development as part of the Alcor program\footnote{\url{https://www.asi.it/en/technologies-and-engineering/micro-and-nanosatellites/alcor-program/}, consulted on 12 Jul 2025}, a program dedicated to the development of CubeSat financed by the Italian Space Agency (ASI). It consists of a 6U-XL platform at the center of which is placed a dual-phase Compton polarimeter optimized for measuring the polarization of hard X-rays in the 25-100~keV band.

\subsection{Compton Polarimetry}\label{sec:compton_pol}

Photons interacts with matter through three phenomena: photoelectric effect, Compton scattering, and pair production. Each of these effects have an energy dependent cross section, which makes them dominant in different energy ranges. While the photoelectric effect is dominant up to keV energies and pair production at very high energies, Compton scattering is the dominant interaction process of hard X-rays in CUSP's energy band. We therefore make use of Compton scattering to determine the polarization of the incoming X-ray flux.

Compton scattering is described by the Klein-Nishina cross section \cite{KN_cross_section_paper}, whose expression is:

\begin{equation}\label{eq:KN_cross-section}
	\dv{\sigma}{\Omega}=\frac{r_e^2}{2}\qty(\frac{E'}{E})^2\qty[\frac{E'}{E}+\frac{E}{E'}-2 \sin^2(\theta)\cos^2(\phi)]
\end{equation}

where $r_e\approx 2.8 \cdot 10^{-15}$~m is the classical radius of the electron, E and E' the energies of the incoming and scattered photon, $\theta$ the polar scattering angle, and $\phi$ the angle between the initial polarization vector and the scattering direction.

The $\cos^2(\phi)$ dependency in the cross section implies that the scattering probability depends on the polarization vector of the incoming photon. Indeed, a photon will preferentially scatter perpendicularly to its polarization vector. Measuring the azimuthal scattering direction of the incoming photon, and repeating the procedure for many photons from a given source, one can infer the polarization parameters of the source. The correlation between the scattering and polarization vector directions implies that the azimuthal scattering angle distribution (a.k.a. modulation curve) is modulated for a polarized source. The amplitude of the modulation is linked to the polarization degree (PD) of the source, while the dip of the $\cos^2$ modulation gives the polarization angle (PA). 

A segmented array of scintillators can be used as a Compton polarimeter. As shown in Figure \ref{fig:CUSP_CAD_Compton}, incoming photons will scatter and be absorbed in two different segments of the instrument, allowing to determine their azimuthal scattering direction. A distribution of the azimuthal scattering angle for many photons from a given source will lead to the determination of its polarization parameters.

\subsection{The CUbesat Solar Polarimeter}\label{sec:cusp}

\begin{figure}[h!]
\centering
 \includegraphics[height=.42\textwidth]{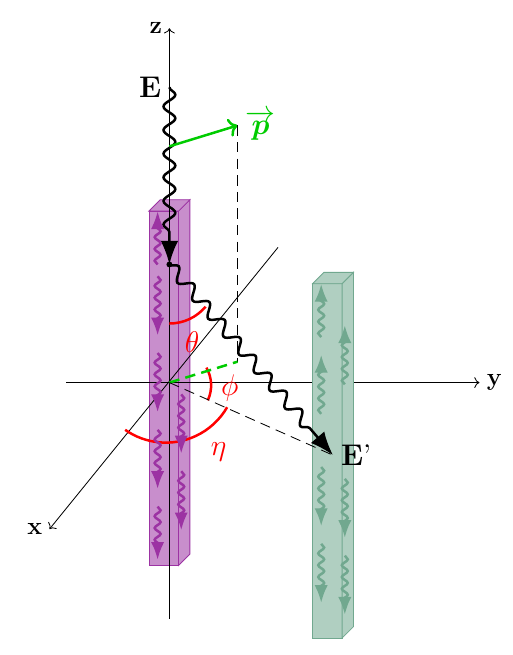}\hspace*{0.5cm}\includegraphics[height=.35\textwidth]{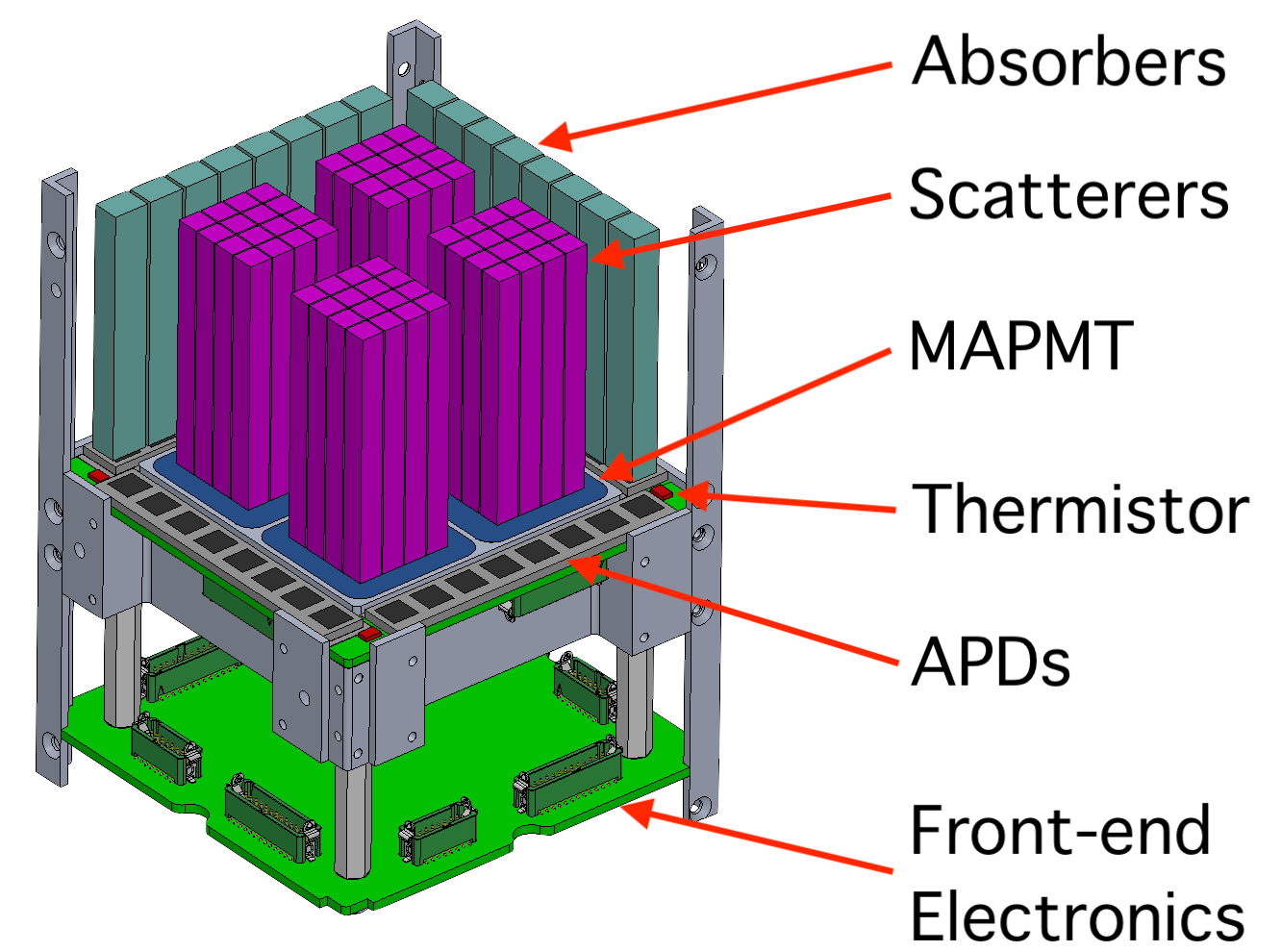}\\[0.5cm]
 \caption{\textbf{Left:} Schematic working principle of a Compton polarimeter. The incoming photon is Compton scattered in a segment of the detector, deposition some energy which is converted into scintillation optical light and collected by a photosensor at the extremity of the scintillator bar. The scattered photon is then absorbed in a different segment of the instrument, which allows to determine the azimuthal scattering direction of the primary photon. \textbf{Right:} CAD design of CUSP's hard X-ray polarimeter sensitive parts.}
 \label{fig:CUSP_CAD_Compton}
\end{figure}

The CUSP mission employs Compton scattering as described in the previous section for measuring the polarization of the non-thermal component of Solar Flares. It consists of a\footnote{The current baseline is to have a single cubesat, but with a possible goal of having a constellation of two cubesats launched in order to increase Sun coverage, polarimetric sensitivity, and reduce systematic effects.} 6U-XL cubesat, with a payload of about 2.5~U. The Compton polarimeter is composed of a central scattering array of 8$\times$8 plastic scintillator bars read out by four 16-channel Multi-Anode PhotoMultiplier Tubes (MAPMTs) from Hamamatsu, surrounded by four strips of 8 GAGG scintillators read out by Avalanche PhotoDiodes (APDs) from Hamamatsu that act as aborbers. MAPMT and APD channels are respectively read out by the MAROC-3A and SKIROC-2A ASICs from Weeroc. A simplified CAD model of the payload showing the relevant components of the polarimeter is depicted in Figure \ref{fig:CUSP_CAD_Compton}.

The 6U-XL cubesat platform has already been developed by IMT s.r.l. for other projects, while CUSP's ground station is located at the University of "La Tuscia" in the north of Latium, Italy. The CUSP project is currently in a 12-month phase B that started in December 2024, at the end of which a polarimeter prototype will be built and tested. Assuming final approval of the mission by ASI — with a key decision expected after the closure of Phase B — the current mission timeline foresees a launch in late 2027 or early 2028.

\section{CUSP PROTOTYPE DEVELOPMENT AND CALIBRATION}\label{sec:proto}

We describe here the first measurements that were performed during phase B using individual channels and ASIC development boards in order to test the functionality and preliminary performances of the instrument.

\subsection{Experimental Setup}\label{sec:single_chan_setup}

\begin{figure}[h!]
\centering
 \includegraphics[height=.4\textwidth]{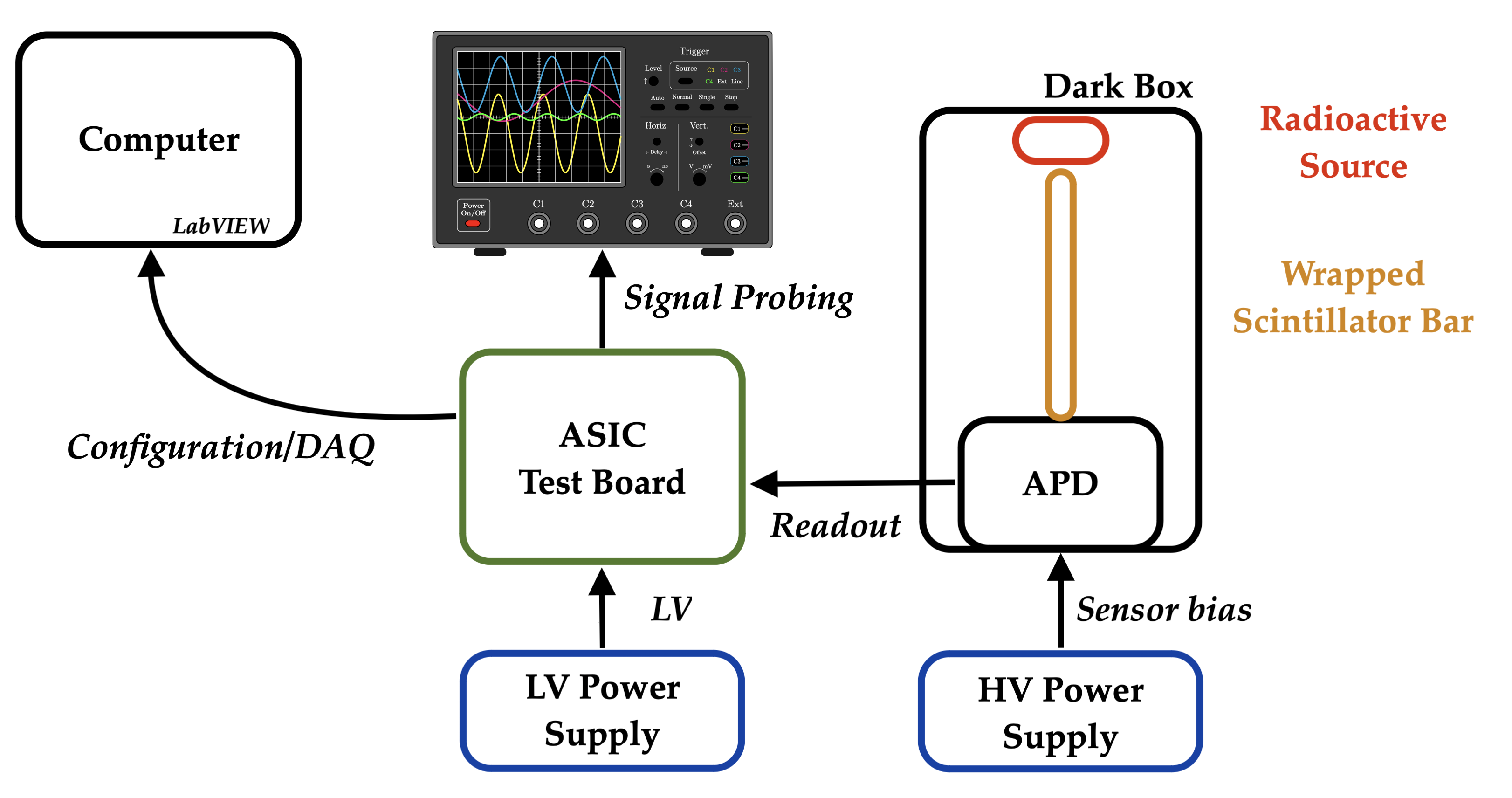}\includegraphics[height=.4\textwidth]{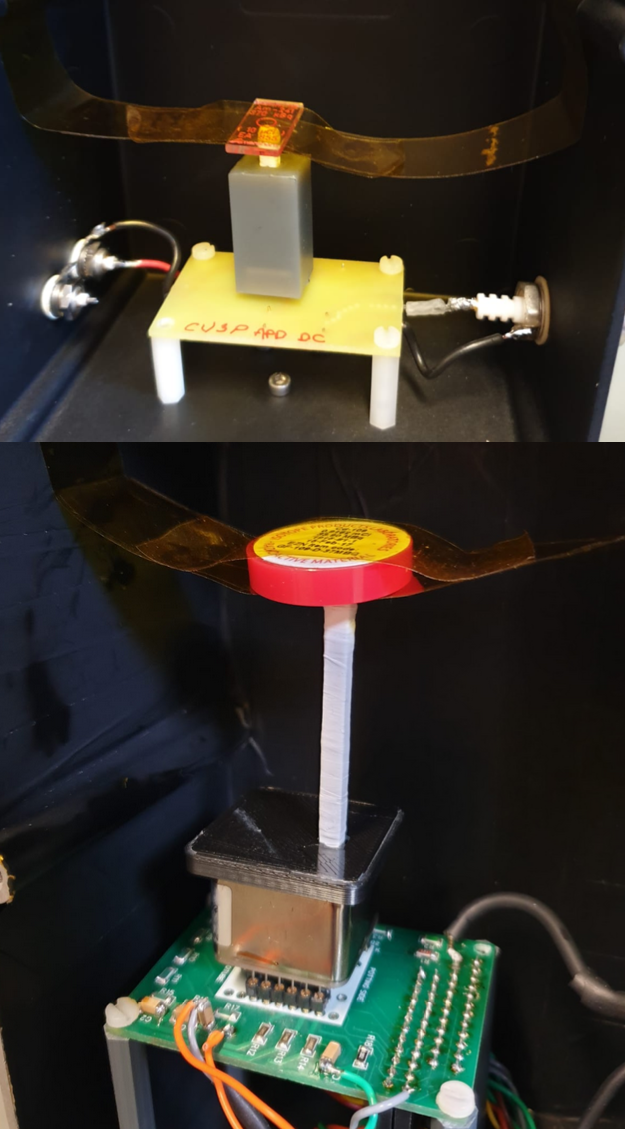}\\[0.2cm]
 \caption{\textbf{Left:} Diagram showing the experimental setup used to test single channels of CUSP's scatterer and absorber using development boards of the flight ASICs. \textbf{Right:} Pictures of a wrapped GAGG bar coupled to an APD with an $^{241}$Am source on top (\textbf{top}) and of a wrapped plastic bar coupled to a MAPMT channel with a $^{109}$Cd source on top (\textbf{bottom}).}
 \label{fig:setup_diagram}
\end{figure}

An experimental setup to test both the scatterer and absorber acquisition chains have been built, as schematically represented in Figure \ref{fig:setup_diagram}.

For the scatterer, a plastic scintillator bar (EJ-204 as a baseline, but other types of plastic scintillators such as EJ-228 and EJ-230 have also been tested) wrapped in PTFE tape is coupled to a channel of a Hamamatsu R7600-03-M16 MAPMT. The MAPMT is mounted on a custom voltage divider board and biased by a Canberra 3005 HV Power Supply NIM module. The MAPMT and scintillator bar are coupled using optical grease, and the assembly is placed in a grounded dark box. The MAPMT is read out through D-sub patch panel and a flat cable by the MAROC-3A evaluation board from Weeroc. The evaluation board is controlled (ASIC configuration, data acquisition) through a dedicated C\# software, while a Teledyne Lecroy Wavesurfer 3104z Oscilloscope is used for probing test signals on the board.

A similar setup has been developed for testing the absorber chain, where a PTFE-wrapped GAGG bar is coupled to an S8664-55 APD from Hamamatsu and the evaluation board of the SKIROC-2A ASIC from Weeroc is used with a dedicated LabView VI for configuration and DAQ.

\newpage
\subsection{Spectral Measurements with Single Channel}\label{sec:single_chan_meas}

\begin{figure}[h!]
\centering
 \includegraphics[height=.38\textwidth]{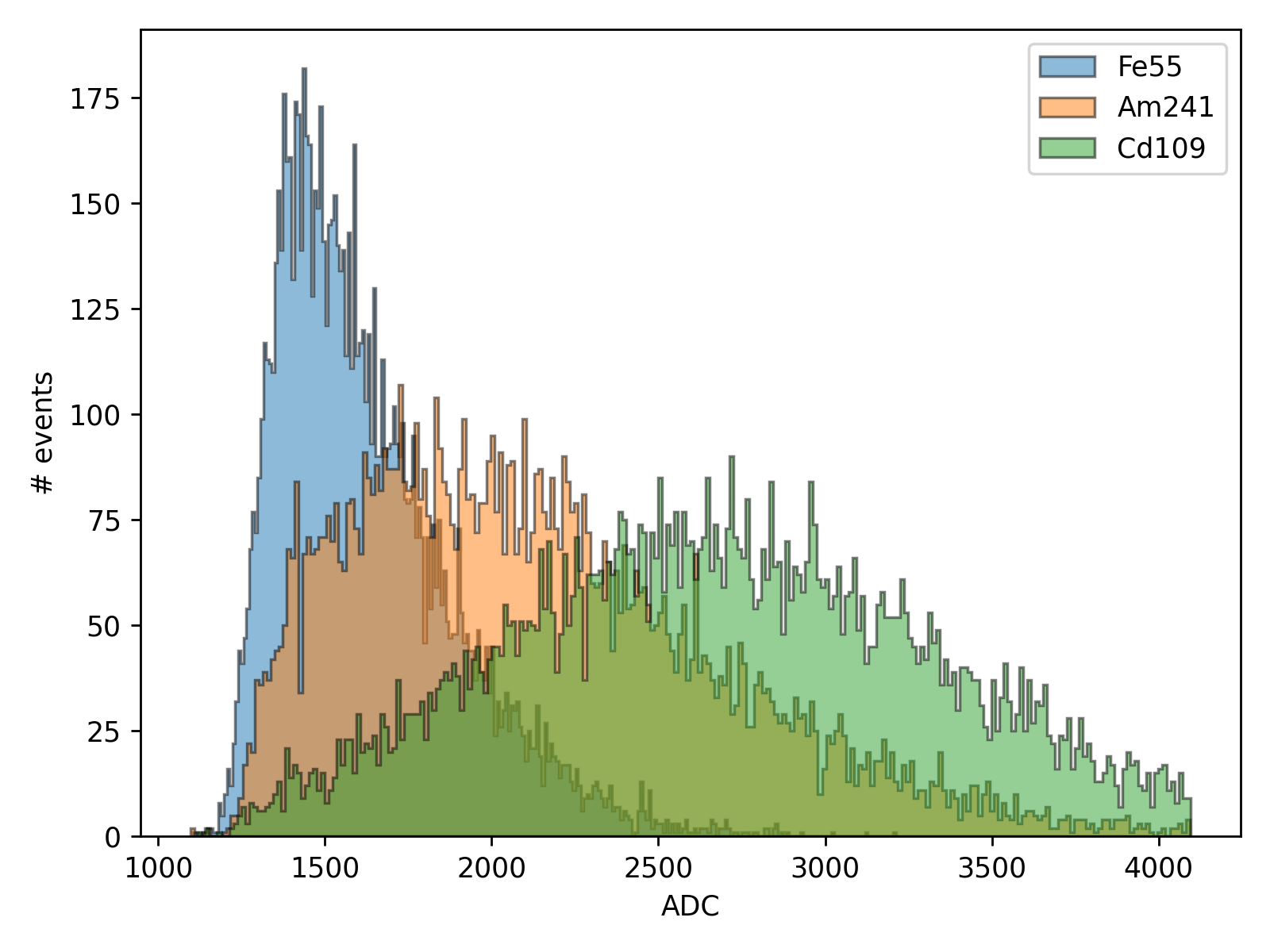}\includegraphics[height=.38\textwidth]{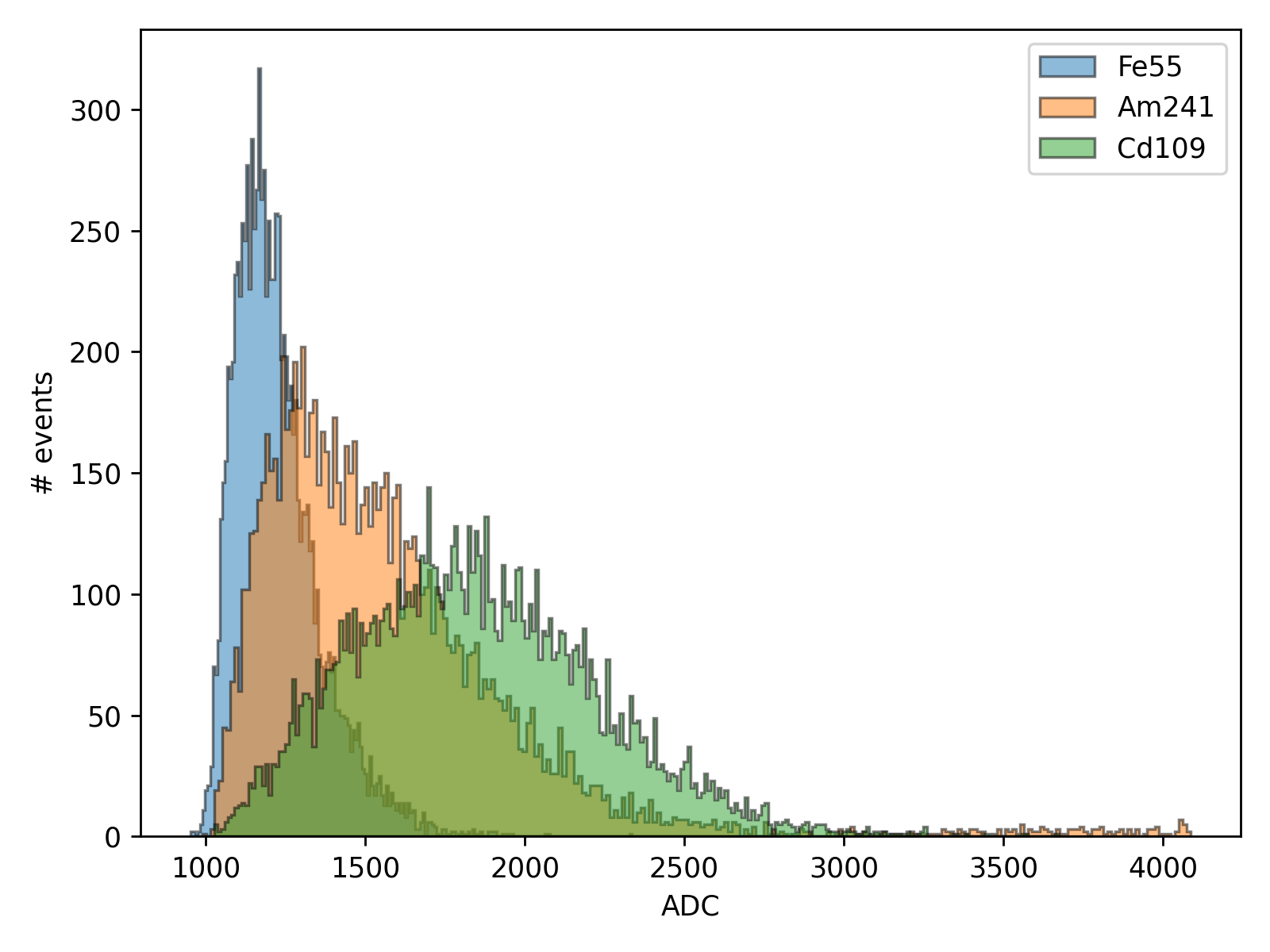}\\[0.2cm]
 \caption{$^{55}$Fe, $^{241}$Am, and $^{109}$Cd spectra acquired using a scatterer channel (EJ-204 plastic+MAPMT) using two types of fast shaper in the ASIC (unipolar fast shaper, \textbf{left}; bipolar fast shaper, \textbf{right}).}
 \label{fig:plastic_spectra}
\end{figure}

Spectra measured using EJ-204 with two different fast shapers (unipolar and bipolar) are shown in Figure \ref{fig:plastic_spectra}. These two internal fast shapers are both being considered for the final configuration of the MAROC-3A ASIC. Lines from the $^{55}$Fe, $^{241}$Am, and $^{109}$Cd sources are visible in both cases. The characteristic energies of these lines are summarized in Table \ref{table:lines}. One should note that the spectra given here are not pedestal subtracted. These measurements will allow to perform an energy calibration of the scatterer channel to show that the required energy range of 1.2-16.4~keV (Compton deposition in the plastic for a 25-100~keV incoming photon in the case of an orthogonal scattering) is achievable.

\renewcommand{\arraystretch}{1.2}
\begin{table}[h!]
\begin{center}
\begin{tabular}{|l|c|c|} 
\hline
 & Lines Detected in EJ-204 [keV] & Lines Detected in GAGG [keV] \\ \hline\hline
$^{55}$Fe & 5.9 & - \\ \hline
$^{241}$Am & 17.7 & 17.7, 59.5 \\ \hline
$^{109}$Cd & 22.6 & 22.6, 88 \\ \hline
$^{57}$Co & - & 122 \\ \hline
$^{155}$Eu & - & 43, 86.5, 105 \\ \hline
$^{129}$I & - & 29.5 \\ \hline
\end{tabular}
\end{center}
\caption{Characteristic energies of the emission lines measured with the plastic and GAGG scintillators for various radioactive source \cite{NuDat}.}\label{table:lines}
\end{table}

\begin{figure}[h!]
\centering
 \includegraphics[height=.42\textwidth]{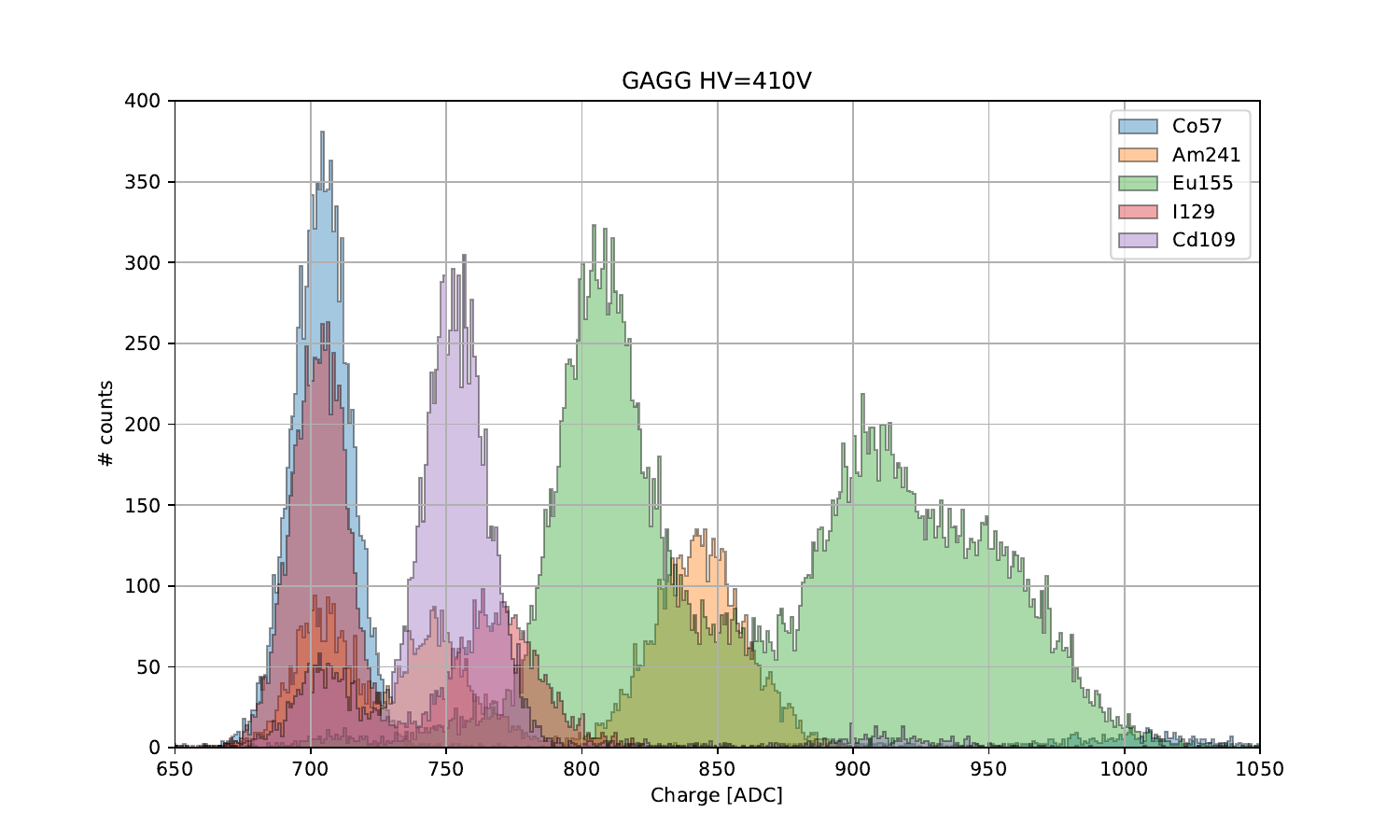}\\[0.2cm]
 \caption{Spectra acquired using an absorber channel (GAGG+APD) with various radioactive isotopes having emission features in CUSP's energy range, namely $^{57}$Co, $^{241}$Am, $^{155}$Eu, $^{129}$I, and $^{109}$Cd.}
 \label{fig:gagg_spectra}
\end{figure}

On the other hand, Americium 241, Cadmium 109, Cobalt 57, Europium 155, and Iodine 129 sources have been used to span the 25 to 100~keV nominal energy range of CUSP with the absorbers. Table \ref{table:lines} once again summarizes the characteristic energies from these sources that are within the energy range of our detector. One can see from the measured spectra in Figure \ref{fig:gagg_spectra}, for a high voltage of 410~V corresponding to an APD gain of about 100, that a good linearity is obtained over the entire energy band. These energy lines will be used to perform an energy calibration of the absorber channels as well as to determine the energy resolution of the system as a function of energy.

\section{CONCLUSIONS \& OUTLOOK}\label{sec:ccl}

The CUbesat Solar Polarimeter, or CUSP, is a Compton polarimetry mission aiming to significantly improve our understanding of heliophysics as well as our prediction capacities for space weather through the measurement of the polarization of the non-thermal component of Solar Flares. It is financed by the Italian Space Agency (ASI) in the frame of its Alcor program for cubesats, and is foreseen for a launch late 2027/early 2028 upon approval of the successive project phases by ASI.

Based on a 6U-XL platform, CUSP will host a 2.5~U Compton polarimeter (including front and back end electronics) made of a central array of plastic scintillator bars surrounded by a frame of GAGG elongated scintillators. We reported here preliminary measurements using both the scatterer acquisition chain based on plastic scintillators coupled to MAPMTs and the absorber chain based on GAGG and APDs. We have shown that we can reach a significant sensitivity over the expected energy range for performing polarimetery in the 25-100~keV band. A more detailed analysis of single channel data is currently ongoing to extract preliminary physical performances of the system \cite{ASAPP_paper}. This will be closely followed by the construction and calibration of a larger scale prototype of the polarimeter made of 16 scatterers and 32 absorbers. This prototype is currently under construction and will be tested by the end of CUSP's phase B (which will end in December 2025). 

The phase B activities will be followed by a proposal for a combined phase C/D to build and test the engineering qualification model and proto-flight model, to pave the way towards the flight model. These later interactions of instruments will be tested in-house at INAF-IAPS using the IXPE Instrument Calibration Equipment \cite{Muleri_ICE}, for which an upgrade is planned to extend the energy band upwards. Possibilities of calibration the instrument on a Synchrotron beamline are also envisaged.

\acknowledgments 
 
This work is funded by the Italian Space Agency (ASI) within the Alcor Program, as part of the development of the CUbesat Solar Polarimeter (CUSP) mission under ASI-INAF contract n. 2023-2-R.0.

\bibliography{BibliographyCUSP} 
\bibliographystyle{spiebib} 

\end{document}